\documentclass[journal]{IEEEtran}

\usepackage{xcolor}
\usepackage{graphicx}
\usepackage{cite}
\usepackage{amsmath,amssymb}
\usepackage{textcomp}
\usepackage[normalem]{ulem}
\usepackage{epstopdf}
\usepackage[export]{adjustbox}
\usepackage{subcaption}
\definecolor{mycolor}{RGB}{0, 0, 0}

\begin{document}



\title{Distortion Mitigation in Millimeter-Wave Interferometric Radar Angular Velocity Estimation Using Signal Response Decomposition}


\author{Eric Klinefelter,~\IEEEmembership{Member,~IEEE}, Jason M. Merlo,~\IEEEmembership{Graduate Student Member,~IEEE}, and \\ Jeffrey A. Nanzer,~\IEEEmembership{Senior Member,~IEEE}

\thanks{The authors are with the Department of Electrical and Computer Engineering, Michigan State University, East Lansing, MI 48824 USA (email: klinefe4@msu.edu, merlojas@msu.edu, nanzer@msu.edu.}
}

\maketitle

\begin{abstract} 
A new method of distortion mitigation for multitarget interferometric angular velocity estimation in millimeter-wave radar is presented. In general, when multiple targets are present, the response of a correlation interferometer is corrupted by intermodulation distortion, making it difficult to estimate individual target angular velocities. We present a distortion mitigation method that works by decomposing the responses at each antenna element into the responses from the individual targets. Data association is performed to match individual target responses at each antenna such that cross-correlation is performed only between associated targets. Thus, the intermodulation distortion (cross-terms) from correlating unlike targets are eliminated, and the result is a frequency response whose individual frequencies are proportional to the angular velocities of the targets. We demonstrate the approach with a custom 40 GHz interferometric radar, a high-accuracy motion capture system which provides ground-truth position measurements, and two robotic platforms. The multitarget experiments consist of three scenarios, designed to represent easy, medium, and difficult cases for the distortion mitigation technique. We show that the reduction in distortion yields angular velocity estimation errors in the three cases of less than $0.008$ rad/s, $0.020$ rad/s, and $0.033$ rad/s for the easy, medium, and hard cases, respectively.
\end{abstract}

\IEEEpeerreviewmaketitle

\section{Introduction}

\IEEEPARstart{M}{ultitarget} tracking refers to the problem of estimating and tracking multiple target states through time by observing noisy measurements \cite{ba_vo}. Measured quantities are typically either kinematic quantities, like position and velocity, or measured attribute quantities, such as signal-to-noise ratio (SNR) or radar cross-section (RCS) \cite{BlackmanSamuelS1999Daao}. Based on the assumption that noise is uncorrelated between measured quantities, the more states we measure or observe, the better tracking performance will be. Traditional radar systems perform three fundamental types of kinematic measurements: range, radial velocity (or Doppler), and angle. Angular velocity may be calculated from sequential angle measurements, but its noise is correlated with the angle measurements and so provides no additional information, and thus no tracking performance improvement by including it in the measurement vector. Our aim with this work is to further explore a fourth basic type of radar measurement, which is interferometric radar angular velocity estimation~\cite{5634150}. Direct angular velocity measurement is independent from the radar angle measurements, allowing estimation of multiple parameters of target motion~\cite{7485234}, and thus can be used to improve radar tracking performance. 


A challenge associated with multitarget tracking in interferometric angular velocity estimation is that of dealing with the non-linear distortion which occurs when multiple targets are present. The ideal interferometric response would produce only the mixing product between like targets in each antenna element response. Instead, the desired mixing products are generated, but the mixing products between unlike targets are generated as well, resulting in intermodulation distortion, or cross-terms, which corrupt the ideal signals. The cross-terms are proportional to both the difference in Doppler and the angular velocities between the two targets, and make recovering the ideal signals difficult.

Interferometric angular velocity distortion was first discussed in~\cite{6888581}, where two hardware-based mitigation methods were presented. One method was to use long-wavelength signals, thus driving the distortion terms, which are proportional to the differential Doppler between targets, to zero. The other method described was to use a pulsed system to separate the targets temporally, and thus perform the pairwise correlation only on like targets. In~\cite{merlo2021multiple} a method using dual-baseline interferometry to reduce the intermodulation distortion was presented, where the normalized time-frequency responses from different baselines are multiplied in order to attenuate the distortion terms, which are scaled at different rates for different baselines. In~\cite{pwang} a similar method was proposed which used the conjugate sum of multiple baselines from a uniform linear array. While these methods yield promising results, they nonetheless require additional hardware resources, either via larger systems, transceivers supporting wide signal bandwidths, or additional receivers. 

In this work we introduce an interferometric distortion mitigation method that is based on decomposition of the individual components of the received signals, and does not require additional hardware. The approach operates either on the two individually received antenna responses, to arrive at angular velocity estimates. The aim of this work is to provide a foundation for signal processing based methods for mitigating interferometric intermodulation distortion terms, and thus accurately estimating the angular velocities of multiple targets. 
The rest of this paper is organized as follows. In Section \ref{ang_vel} we review interferometric angular velocity estimation and provide the general model for the interferometric response to an arbitrary scene containing $N$ point scatterers. In Section \ref{dist_mit} we introduce the two classes of signal processing based distortion mitigation, and make a case for pursuing the class of methods based on response decomposition. Finally, in Section \ref{experiments} we present our experimental measurements with a 40 GHz radar that provides both angular velocity and radial velocity measurements. We investigate the feasibility of the approach through the estimation of the angular and radial velocities of moving robotic platforms.


\section{Interferometric Angular Velocity Estimation} \label{ang_vel}

\begin{figure}[t!]
\centering
\includegraphics[width=1.0\linewidth]{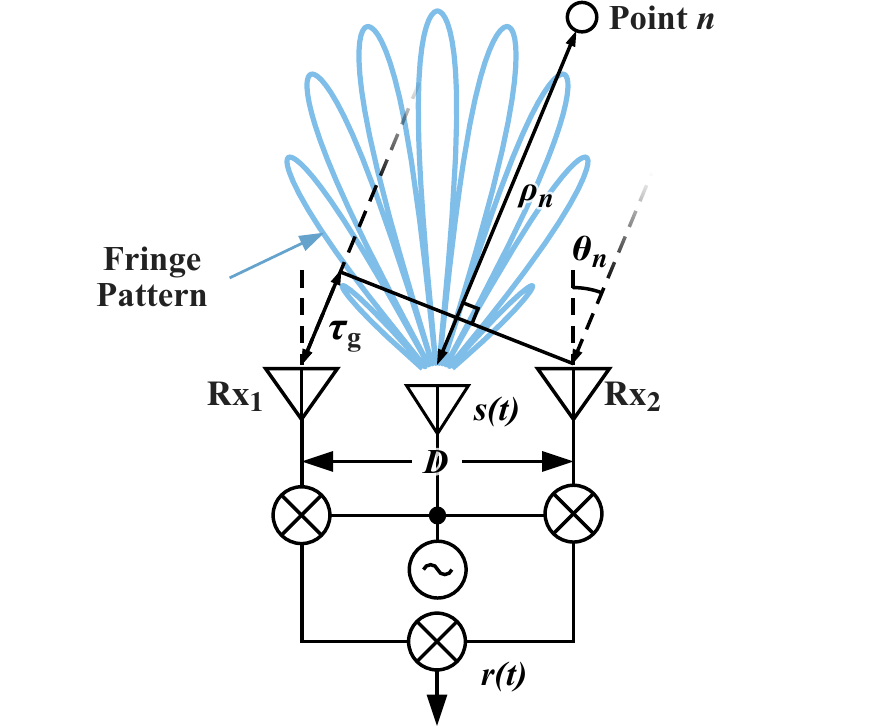}
\caption{Block diagram of a two-element correlation interferometer with system baseline $D$, transmitted signal $s(t)$, and correlator output $r(t)$. The fringe pattern creates a series of peaks and nulls created by the constructive and destructive interference of the received signals at various angles. The geometric time-delay of a plane wave is shown as $\tau_g$ and assumes that the target is far-field to the array, which implies $\theta_{1,n} \approx \theta_{2,n}$.}
\label{interf_schem}
\end{figure}

Direct measurement of the angular velocity is enabled by cross-correlating the received signals in a two-element interferometric receiver; the object may be moving~\cite{5634150}, or the interferometer may be rotating, as in radio astronomy~\cite{thompson2017interferometry}. A diagram of a two-element correlation interferometer and its associated fringe pattern resulting from the cross-correlation process is shown in Fig. \ref{interf_schem}. As a target passes through the fringe pattern in angle, an oscillation is observed, where the period of the oscillation is inversely proportional to the angular velocity of the target. Observing a point target in the far-field, the instantaneous frequency of the output of a correlation interferometer is \cite{5634150}
\begin{align} \label{shift}
f_i = \frac{\omega D}{{\lambda_c}} \cos (\omega t),
\end{align}
where $D$ is the system baseline, or antenna separation, $\lambda_c$ is the transmitted wavelength, and $\omega$ is the angular velocity of the target. For angles close to broadside where \mbox{$\cos(\theta) \approx 1$}, the instantaneous frequency is directly proportional to the target's angular velocity. When two targets are present there are two frequency shifts as determined by (\ref{shift}), which are the result of correlating each target with itself. There are also two additional components, or cross-terms, which are the result of correlating target one with target two and vice versa. These additional terms are not realted to the angular velocity of a real target, and thus distort the ideal interferometric responses.

In \cite{npoint} we derived the exact interferometric response to a target containing $N$ ideal point scatterers. We also provided an approximate response which is valid when we make the far-field and small-angle approximations. The approximate spectral response is
\begin{align} \label{tot_spect}
&R(f) = \sum\limits_{n=1}^{N} \sum\limits_{k=1}^{N}  A(\theta_{1,n})A^*(\theta_{2,k}) \delta \left( f-f_{n,k} \right),
\end{align}
where $N$ is the number of point targets, $A$ is the complex antenna voltage pattern, and $\theta_{i,n}$ is the angle of target $n$ referenced to antenna $i$. For each target pair $\lbrace n,k \rbrace $, the resulting frequency shift is
\begin{align}\label{components}
f_{n,k} =  \frac{2 ( v_{r,1,n}-v_{r,1,k})}{\lambda_c} + \frac{\omega_{1,k} D}{ \lambda_c},
\end{align}
where $v_{r,1,n}$ is the radial velocity of target $n$ referenced to \mbox{antenna one} (RX1), and $\omega_{1,k}$ is the angular velocity of target $k$ referenced to antenna one. In the case where $n=k$ we get the single point response given in (\ref{shift}) with the small-angle approximation. We consider this the ideal interferometric response for that point target. When $n\neq k$ we see that we get a frequency shift value that is proportional to not only the angular velocity of target $k$, but also the difference in the radial velocities between the two points. These are the cross-terms which corrupt the ideal responses.

\section{Distortion Mitigation Approaches} \label{dist_mit}

Due to the double summation in (\ref{tot_spect}), the interferometric response of a scene containing $N$ scatterers will contain $N^2$ frequency components. $N$ of these components will contain angular velocity information, which occurs when a point response is correlated with itself. There will also be $N(N-1)$ distortion terms, which occurs when each point is correlated with every other point. The goal of  distortion mitigation is to recover the angular velocity information of each target. But since the closed-form expression for the response is not one-to-one, these terms cannot be directly obtained, i.e., two scenes each with different targets having different angular and radial velocities can generate the same spectral response. In this section we discuss two main classes of techniques for solving this problem. The first uses a model-based response with optimization techniques to estimate the ideal underlying angular velocity parameters, while the second approach, response decomposition, involves using methods aimed at avoiding generating the $N(N-1)$ distortion terms.

\subsection{Model-Based Methods}


The problem of distortion mitigation is simplified in the case where a closed-form representation is available, as given in (\ref{tot_spect}). The response is essentially a parameterized non-linear function for which we would like to find the unknown parameter (angular velocity) values. We do this by finding parameter values which minimize the error between the observed response and the calculated model response with the estimated parameters. A brute-force approach would be to try every possible combination of angular velocity values in order to find the model response which perfectly matches the measurement. In practice, more efficient optimization techniques are available, such as gradient descent, which is guaranteed to approach the true parameter values when the loss space is convex \cite{363438}. 
The challenge with this method is that, in practice, the loss space is non-convex. This can be mitigated to some extent by performing a coarse grid search to find initial parameter seed values that may find the convex part of the loss space that contains the true parameter values \cite{Du2019}. Another difficulty with this method, and any method that operates on the complete interferometric response, is that the number of frequency components of the response is $N^2$, thus as the number of targets grows the number of components grows quadratically, making for highly non-convex loss spaces with many local minima.

Model-based methods could be used to mitigate interferometric distortion, even if the model in (\ref{tot_spect}) were not available, by applying learning methods to acquired data,
which may be achieved using feedforward or recurrent neural networks \cite{Goodfellow-et-al-2016}. The challenge with this approach is the same as for all learning-based approaches: a large amount of training data is required, especially for use with deep neural networks \cite{ijcai2017-318}. This is simple to do in simulation where direct access to the individual target responses are readily available, from which the pairwise correlations can be calculated to arrive at the ideal interferometric response. This response then contains no distortion and is used as the target for the neural network. The inputs to the network may be the amplitude and phase values of the distorted interferometric response and even the individual Doppler responses if desired. For learning to occur the spectrum should not be represented with delta functions, which will have substantial loss function gradients only at discrete intervals \cite{werbos1990backpropagation}. However, in measured data it is generally challenging to determine the ideal interferometric responses, thus to train on actual measurements the distortion would either need to be hand filtered or filtered using another distortion mitigation.

Due to the challenges posed by the two approaches above, we seek a distortion mitigation technique that modifies the incoming signals in order to avoid generating the $N(N-1)$ distortion terms in the first place. This approach is discussed in the next section.

\subsection{Response Decomposition Methods}

Distortion terms arise from the simultaneous presence of signals from multiple targets in the correlator. Response decomposition methods aim to separate the incoming signals into their respective components such that digital cross-correlation can be performed on only the responses from the same targets, thus eliminating the generation of cross-terms.
Earlier we mentioned a hardware-based distortion mitigation technique in which a pulsed system could be used to temporally isolate the responses from individual targets \cite{6888581}. This could be considered a hardware based version of response decomposition that operates by separating the responses from multiple signals in the time domain. Here we take a related approach, but one that operates solely in signal processing, and that isolates the individual responses from each target in each antenna in the frequency domain.


The method of distortion mitigation using response decomposition seeks to ensure that the number of targets in (\ref{tot_spect}) is $N=1$, i.e. only one target response is present, representing the case when $n=k$. The response at antenna $i$ due to a scene with $N$ independent point scatterers is
\begin{align}\label{ant1}
s_i(t) &= \sum\limits_{n=1}^N s_{i,n}(t),
\end{align}
and the individual response due to a single target $n$ is
\begin{align}
s_{i,n}(t) &= A(\theta_{i,n}) e^{j 2 \pi f_c (t - \tau_{i,n}(t)) },
\end{align}
where $f_c$ is the transmitted center frequency, and $\tau_{i,n}$ is the time-delay due to the distance traveled from the transmitter to point $n$ and back to antenna $i$. A complex correlation interferometer performs conjugate multiplication
\begin{align}
r(t) &= \langle s_1(t)s_2^*(t) \rangle ,
\end{align}
where $\langle \cdot \rangle$ represents time-averaging or low-pass filtering. This process will result in the spectral output given in (\ref{tot_spect}). The response decomposition method decomposes the measured signals at antenna one, $s_1(t)$, and antenna two, $s_2(t)$, into their individual $N$ terms as described in (\ref{ant1}). It also assumes that after decomposing each response we are able to associate the terms with their correct counterpart in the other antenna response. For example, if each antenna response contains two terms we must know which terms were generated by the same target, and match those two. Thus the output of the correlation interferometer using response decomposition is
\begin{align}
r_D(t) &=  \sum\limits_{n=1}^N \langle s_{1,n}(t)s_{2,n}^*(t) \rangle .
\end{align}
Here we see that the total response is a summation of the correlations of the responses at each antenna only due to the individual points $n$. Thus the total spectral response is
\begin{align} \label{tot_spect_DD}
R_D(f) = \sum\limits_{n=1}^{N}  |A(\theta_{1,n})|^2 \delta \left( f-\frac{\omega_{1,n} D}{\lambda_c}  \right) .
\end{align}
From this we see that the total response contains frequency shifts which are directly proportional to each target's angular velocity. 

Time-domain response decomposition is possible for short-pulse systems, however, for continuous-wave systems, decomposition is more easily accomplished in the frequency domain. The spectral responses of the received signals at antenna one and two are $S_1(f)$ and $S_2(f)$, from which it is desired to generate $R_D(f)$. Thus, we first decompose each antenna's spectral response into components due to each target as
\begin{align}
S_i(f) &=  \sum\limits_{n=1}^N S_{i,n}(f),
\end{align}
where $i$ is the antenna element. Conjugate multiplication is equivalent to convolution in the frequency domain, thus we generate the ideal response for a specific target $n$, by convolving the spectral responses from each antenna element as
\begin{align}
R_{D,n}(f) &=  S_{1,n}(f) \star S_{2,n}(f).
\end{align}
Thus the total spectral response using response decomposition is the sum of the individual ideal responses as
\begin{align}
R_D(f) &=  \sum\limits_{n=1}^N R_{D,n}(f) .
\end{align}
The ability to generate this response depends on how well the individual responses can be isolated, i.e., how well the signal can be decomposed into the components corresponding to separate targets, and also on how well the response from like targets can be associated between the two antennas.

%

\subsubsection{Signal Decomposition}
For ideal point targets, the process of detecting a target in the response of an individual antenna is equivalent to sinusoidal parameter estimation in the presence of noise. An approximation to the maximum likelihood estimator (MLE) for the frequency parameter is the maximum of the periodogram \cite{Kay1993}. Thus, one simple method for signal decomposition may consist of detecting peaks above some threshold in each antenna's spectral response. If the number of targets is known a priori, super-resolution techniques may be used, such as MUSIC or ESPIRIT to obtain higher resolution frequency estimates \cite{1143830, 32276}. In general though, the number of targets may be both unknown and time-varying, thus the number of targets present needs to be estimated. Furthermore, due to noise and limited frequency resolution there may be cases where the cardinality of targets is not equivalent between the antenna responses. Also, in some cases, as in the measurements shown later, a target response is not easily modeled as an ideal sinusoid. Instead of simple peak detection, we use a metric based on window locations with maximum integrated power to detect the target locations in the frequency domain. Regardless of the method used, it is critical to detect,  locate and isolate the portion of the frequency spectrum that is attributable to a specific target. This may be accomplished through masking, bandpass filtering, or similar methods.

\subsubsection{Data Association}
In traditional data association for multitarget tracking, targets are associated by computing the likelihood that two responses are associated with each other based on some a priori knowledge. For example, if range is the parameter being tracked an expected motion model exists, the most likely location at the next time step can be computed based on the target's velocity. A Gaussian probability density function (PDF) centered at this location can then be defined, and the PDF at the value of the new measurement can be evaluated to find the probability that the new measurement belongs to the previous target. The process is very similar for associating detections between the responses at each antenna of the correlation interferometer, where frequency is being tracked, which is proportional to the target's angular velocity. 
A simple and effective data association method is the global nearest neighbor (GNN) approach, which associates targets which minimize the total distance (frequency separation) between the targets.

The problems of signal decomposition and data association are not trivial, and have been studied in the literature \cite{Skolnik2001, Levanon2004, BarShalom1995MultitargetMultisensorTP}. Since these problems are well defined and response decomposition mitigates the quadratic nature of the growing distortion terms, this method is preferable over model-based techniques described earlier. In the next section we demonstrate this method on millimeter-wave radar measurements.

\section{Experimental Evaluation} \label{experiments}

\subsection{Measurement System}

\begin{figure}[t]
\centering
\includegraphics[width=1.0\linewidth]{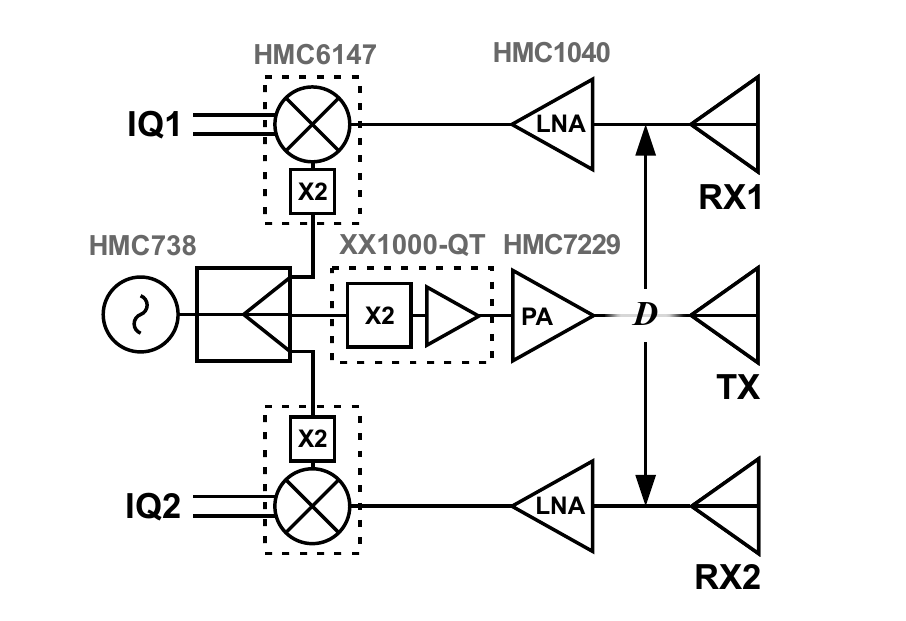}
\caption{Schematic of millimeter-wave radar used for experimental measurements. System consists of two phase-quadrature direct-downconversion receive channels with a continuous-wave transmitter operating at 40 GHz.
}
\label{radar_schem}
\end{figure}

\begin{figure}[t]
\centering
\includegraphics[width=0.98\linewidth]{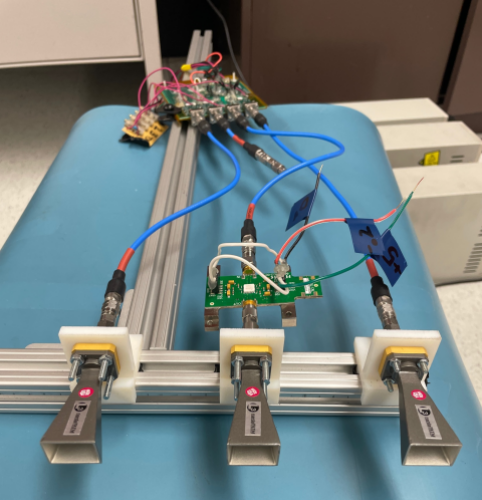}
\caption{The 40 GHz interferometric radar consisted of a custom single board radar and L3-NARDA 15 dBi horn antennas. The transmitter was located in the center between the two receivers which were separated by $20\lambda$.
}
\label{radar}
\end{figure}

We evaluated the response decomposition method using a \mbox{40 GHz} two-element continuous-wave (CW) interferometric radar, shown schematically in Fig. \ref{radar_schem}.
The system was built using commercially available RF monolithic microwave integrated circuit (MMIC) components, and the \mbox{40 GHz} center frequency was chosen as a tradeoff between short wavelengths to support a small physical baseline and the cost and availability of ICs. The PCB was designed on Rogers 4350 0.020" substrate using Ansys HFSS to desing the transmission lines and Wilkinson power splitters \cite{pozar2011microwave}. The design consists of a 20 GHz HMC738 voltage-controlled oscillator (VCO), which serves as the CW transmit signal and the local-oscillator (LO) for the IQ downconverters. The signal is split using a 1:4 Wilkinson power splitter and fed to an XX1000-QT frequency doubler before being amplified by an HMC7229 power amplifier (PA). The received signals are amplified by an HMC1040 low-noise amplifier (LNA) before being directly downconverted with an HMC6147 IQ downconverter. The downconverters have built in LO x2 frequency multipliers. The baseband in-phase and quadrature signals are amplified with inverting op-amps and sampled with an NI USB-6002 DAQ. The antennas used in this work were L3-NARDA 15 dBi standard gain horn antennas, positioned to achieve a $20\lambda$ interferometric baseline. An image of the measurement system is shown in Fig. \ref{radar}.

\begin{figure}[t]
\centering
\includegraphics[width=1.0\linewidth]{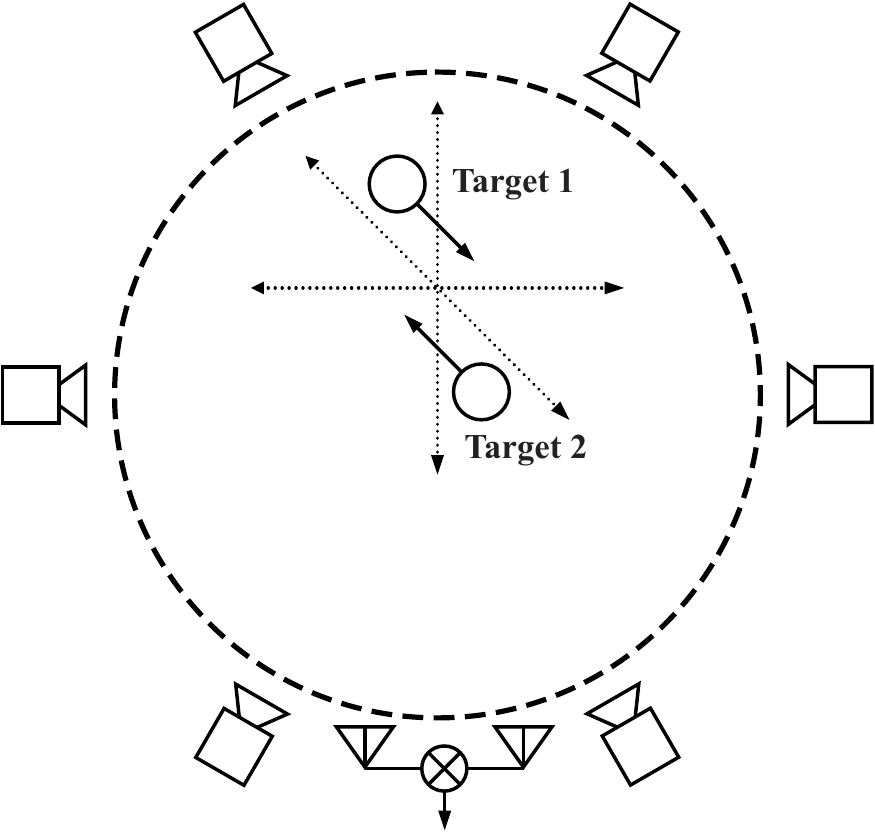}
\caption{Top-down view of the experimental setup. The six motion capture cameras form the capture area shown as the dashed circle. The radar is placed at the edge of the area and the origin was set to the center of the interferometric array. The targets traveled in various trajectories as shown.
}
\label{experiment}
\end{figure}

\begin{figure}[t]
\centering
\includegraphics[width=0.98\linewidth]{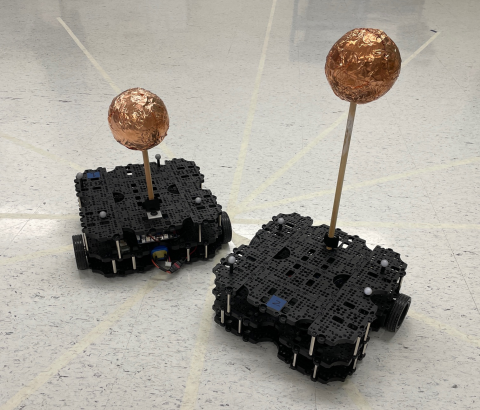}
\caption{Robotis Turtlebot3 Waffle platforms used for measurements. The reflective targets consisted of polystyrene spheres coated in copper tape.
}
\label{turtlebots}
\end{figure}

To obtain ground-truth target position measurements an OptiTrack motion capture system was used. This system comprised six Flex 13 infrared cameras with a 120 Hz frame rate and stated position accuracy of 0.2 mm. A top-down diagram of the experimental setup is shown in Fig. \ref{experiment}. The six cameras are mounted from the ceiling and create the capture area shown as the dashed circle. The origin is set as the center of the interferometric array, and the targets travel in various trajectories as shown. Ground-truth radial and angular velocity values were computed as the time-rate change of range and angle measurements. The expected Doppler and interferometric frequency shifts were computed at each time instance using the relationship between the Doppler shift and radial velocity, $f_D=2v_r/\lambda$, and the expression for the interferometric shift in (\ref{shift}). The targets consisted of two Robotis Turtlebot3 Waffle platforms with reflective targets made of polystyrene spheres coated in copper tape, as shown in Fig. \ref{turtlebots}.

\subsection{Single-Target Experiments}

\begin{figure*}[t]
\centering
\begin{subfigure}{1.0\textwidth}
\includegraphics[width=\linewidth]{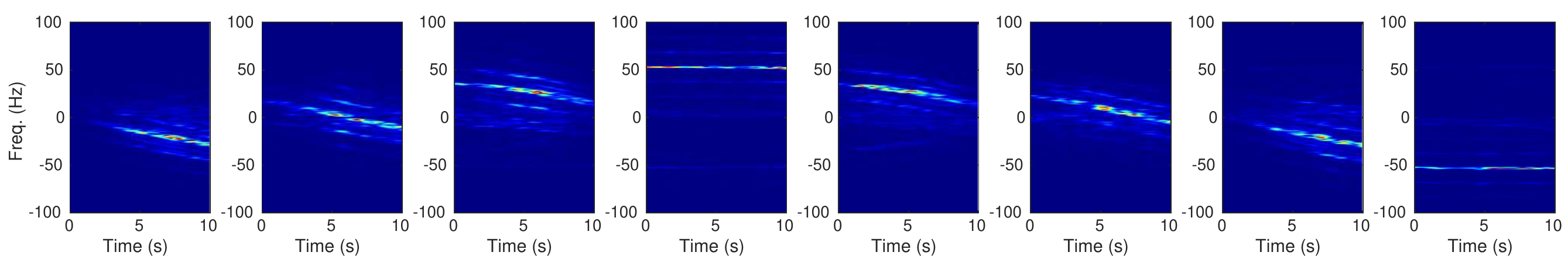}
\caption{}
\end{subfigure}
\begin{subfigure}{1.0\textwidth}
\includegraphics[width=\linewidth]{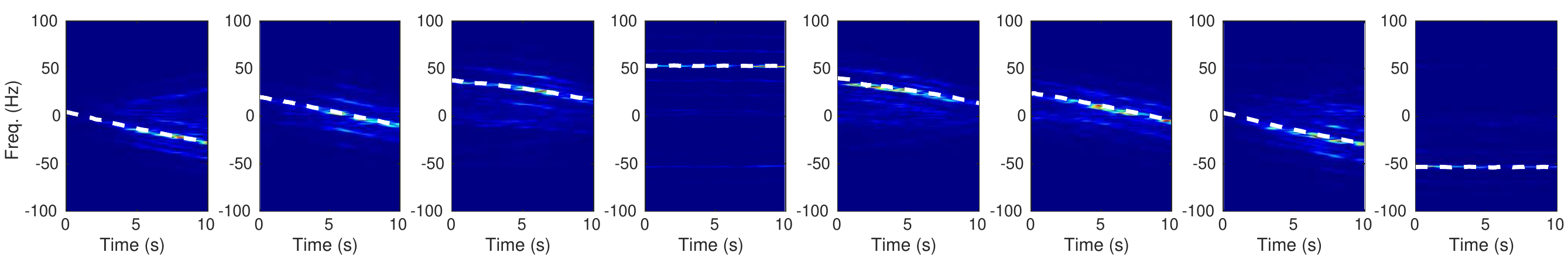}
\caption{}
\end{subfigure}
\begin{subfigure}{1.0\textwidth}
\includegraphics[width=\linewidth]{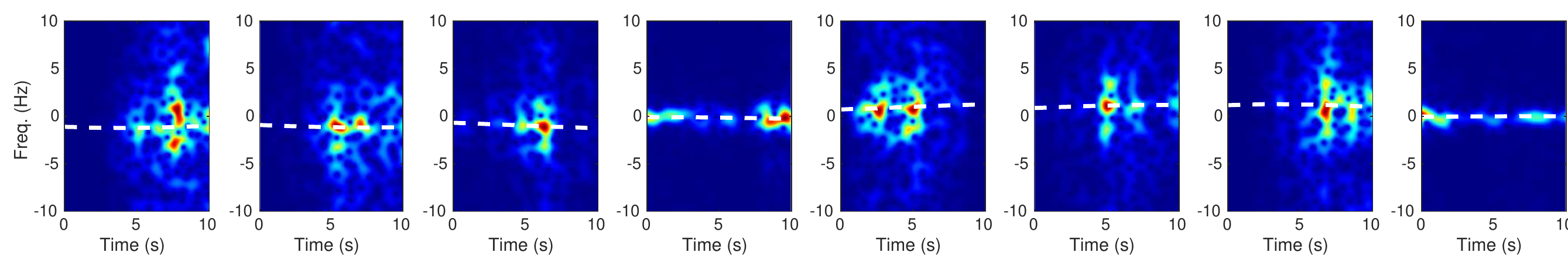}
\caption{}
\end{subfigure}
\caption{The time-frequency responses for a single target at various trajectory angles shown in each column. The plots are the Doppler response from antenna one (a), the Doppler response from antenna two with the expected Doppler shift overlaid in a white dashed line (b), and the interferometric response with the expected interferometric shift (c). The Doppler responses of columns two and six look identical, but we can notice a small negative interferometric shift in column two, and a small positive shift in column six, implying counter-clockwise and clockwise motion. 
}
\label{single_tf}
\end{figure*}

To validate the radar measurement and ground-truth position system we first performed a number of single target measurements with the target traveling at angles from $0-360^\circ$ relative to the radar. The in-phase and quadrature radar signals were captured at 1.92 kHz with the DAQ and high-pass filtered with a 3rd order Butterworth filter with a cutoff of 1 Hz to remove any DC bias. The short-time Fourier transform (STFT) was performed on each individual antenna response to generate the antenna one and antenna two Doppler responses. A window length and FFT size of 1024 samples was used with an overlap of 960 samples. The interferometric response was generated by conjugate multiplication in the time-domain of the individual antenna responses. The time-frequency responses for various trajectories are shown in Fig. \ref{single_tf}. In each column we see a different target angle trajectory, while the plots from top to bottom are the Doppler response from antenna one (a), the Doppler response from antenna two with the expected Doppler shift in a white dashed line (b), and the interferometric response with the expected interferometric shift in a white dashed line (c). Here we see good agreement between the measured and expected frequency shifts. 

Columns four and eight comprise of purely radial motion where the highest positive and negative Doppler shifts manifest, and the interferometric shifts are zero, implying no angular motion. The strength of the interferometric measurement is highlighted by comparing the remaining columns, where the Doppler responses are similar, but the interferometric responses have a detectable frequency shift. For example, the interferometric responses show a negative frequency shift in column two, indicating angular velocity in the counter-clockwise direction, and a positive frequency shift in column six, indicating angular velocity in the clockwise direction. 

\subsection{Multitarget Experiments}

\begin{figure}[t]
\centering
\begin{subfigure}{0.15\textwidth}
\includegraphics[width=\linewidth]{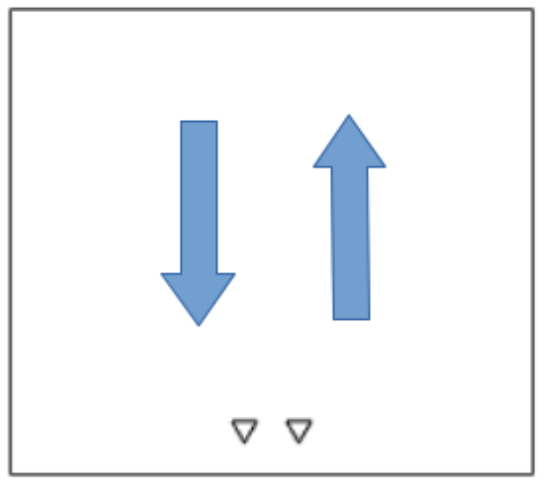}
\caption{}
\end{subfigure}
\begin{subfigure}{0.15\textwidth}
\includegraphics[width=\linewidth]{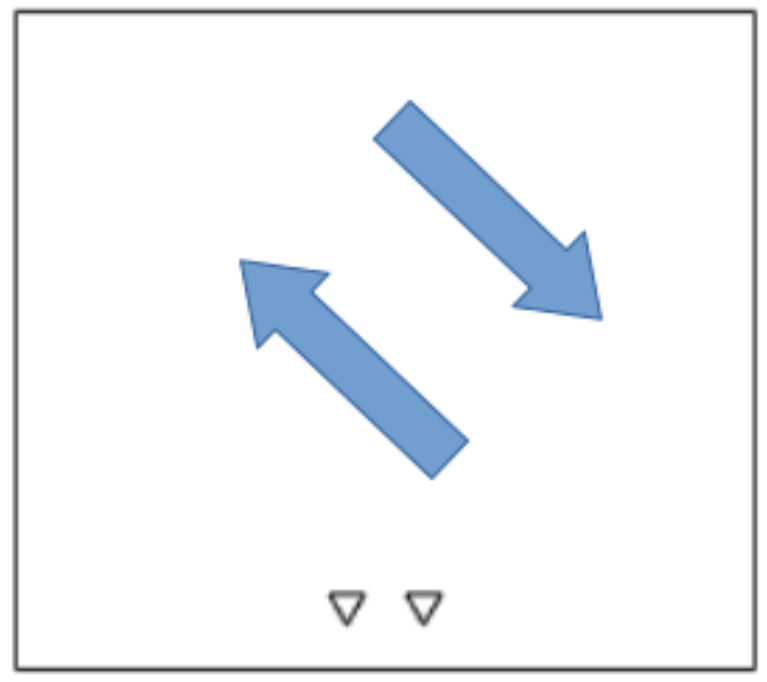}
\caption{}
\end{subfigure}
\begin{subfigure}{0.15\textwidth}
\includegraphics[width=\linewidth]{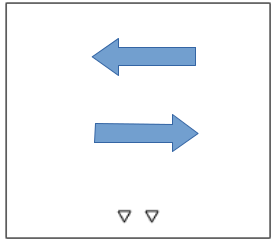}
\caption{}
\end{subfigure}
\caption{Robot trajectory directions for the easy (a), medium (b), and difficult case (c), with the antenna receiving pair shown at the bottom of each image. The large differential in target Doppler shifts will be easily separable in case (a), while the negligible Doppler shifts in case (c) will not be easily separable.
}
\label{traj}
\end{figure}

To test the effectiveness of the distortion mitigation technique we performed measurements with two targets in three distinct motion patterns which were designed to highlight easy, medium, and difficult cases for the distortion mitigation technique, as shown in Fig. \ref{traj}. The trajectories shown in (a) result in positive and negative Doppler frequency shifts, in which isolation of the Doppler responses should be at its easiest. However, in this case there is no angular motion and so the interferometric shifts will be zero for each target. Case (b) shows trajectories which result in both positive and negative Doppler shifts, as well as positive and negative interferometric shifts. Finally, case (c) results in the largest positive and negative interferometric shifts, though the Doppler shifts are both negligible, thus isolating the individual Doppler frequencies for response decomposition should be most challenging.

\begin{figure}[t]
\centering
\begin{subfigure}{0.5\textwidth}
\includegraphics[width=\linewidth]{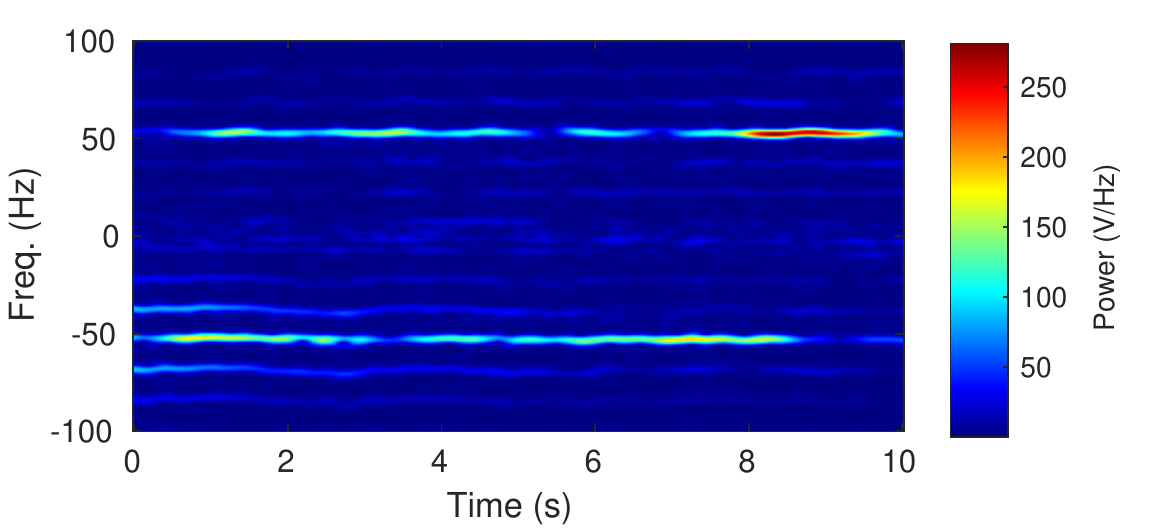}
\caption{}
\end{subfigure}
\begin{subfigure}{0.5\textwidth}
\includegraphics[width=\linewidth]{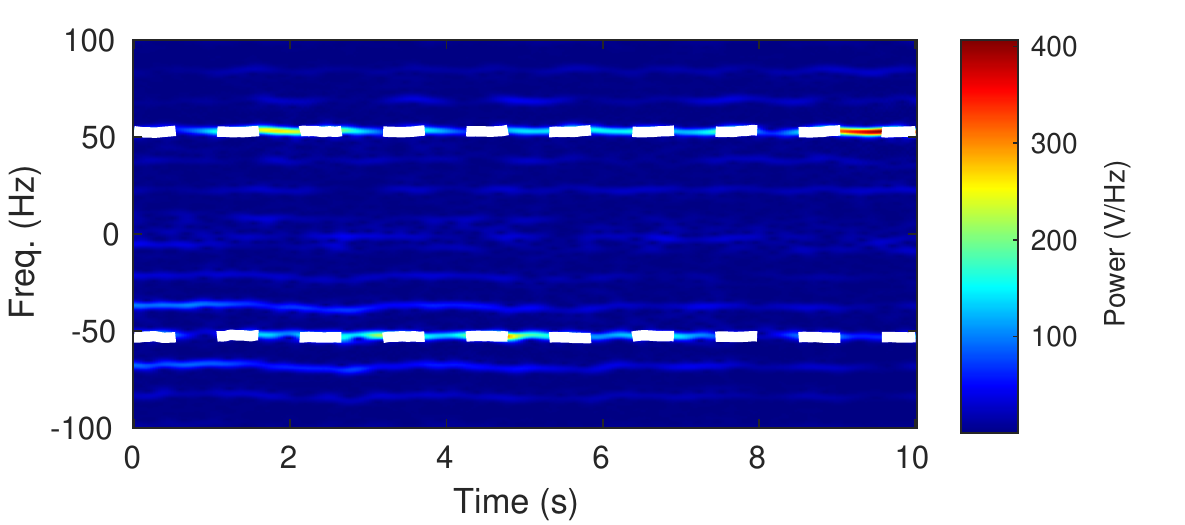}
\caption{}
\end{subfigure}
\begin{subfigure}{0.5\textwidth}
\includegraphics[width=\linewidth]{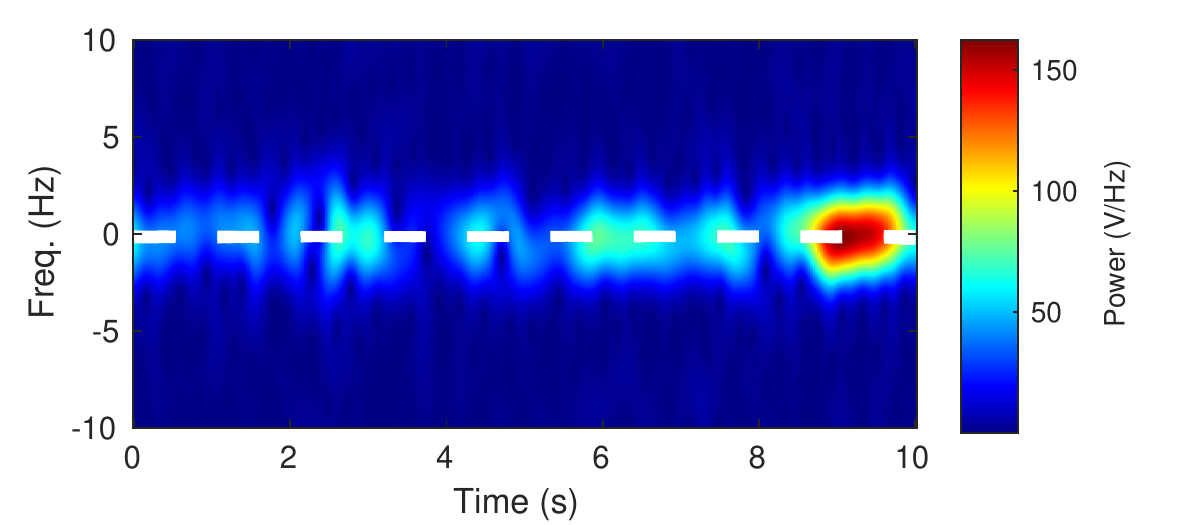}
\caption{}
\end{subfigure}
\caption{The measured time-frequency responses for the easy case include the antenna one Doppler response (a), antenna two Doppler response with ground-truth expected Doppler shift (b), and interferometric response with the expected interferometric shift (c).
}
\label{r37}
\end{figure}

\begin{figure}[t]
\centering
\begin{subfigure}{0.5\textwidth}
\includegraphics[width=\linewidth]{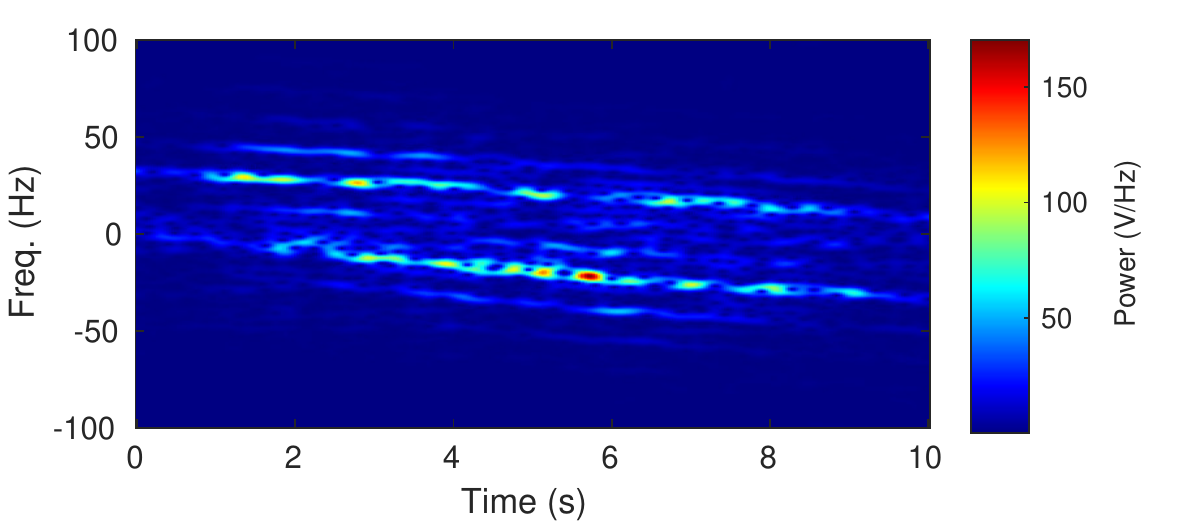}
\caption{}
\end{subfigure}
\begin{subfigure}{0.5\textwidth}
\includegraphics[width=\linewidth]{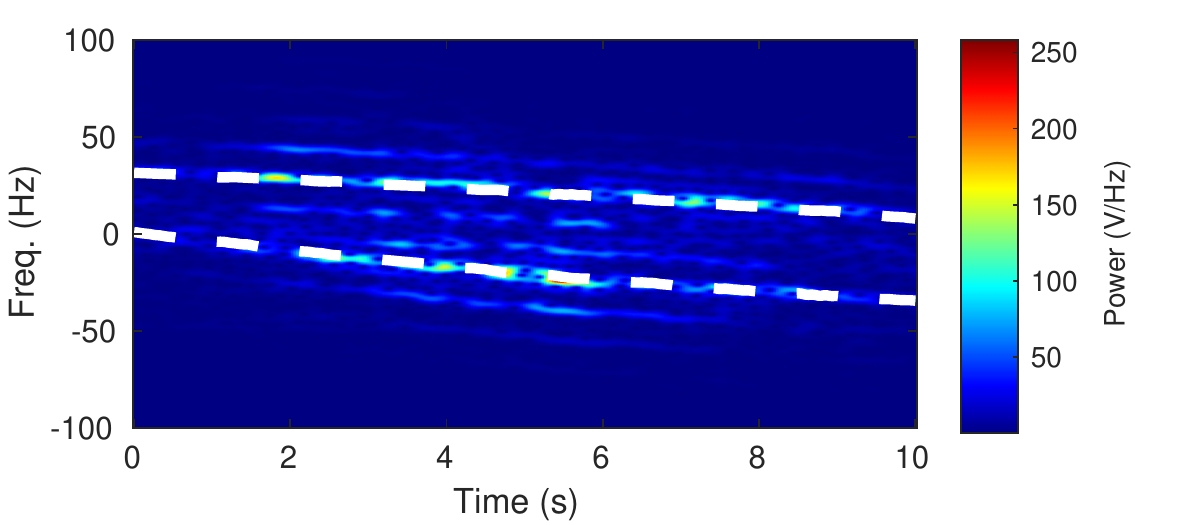}
\caption{}
\end{subfigure}
\begin{subfigure}{0.5\textwidth}
\includegraphics[width=\linewidth]{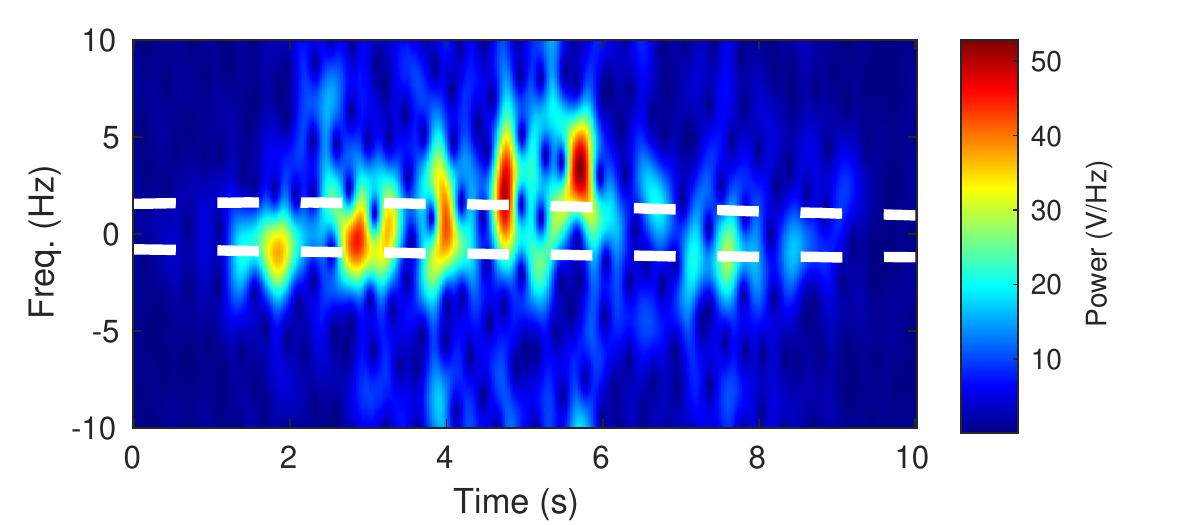}
\caption{}
\end{subfigure}
\caption{The measured time-frequency responses for the medium case include the antenna one Doppler response (a), antenna two Doppler response with ground-truth expected Doppler shift (b), and interferometric response with the expected interferometric shift (c).
}
\label{r40}
\end{figure}

\begin{figure}[t]
\centering
\begin{subfigure}{0.5\textwidth}
\includegraphics[width=\linewidth]{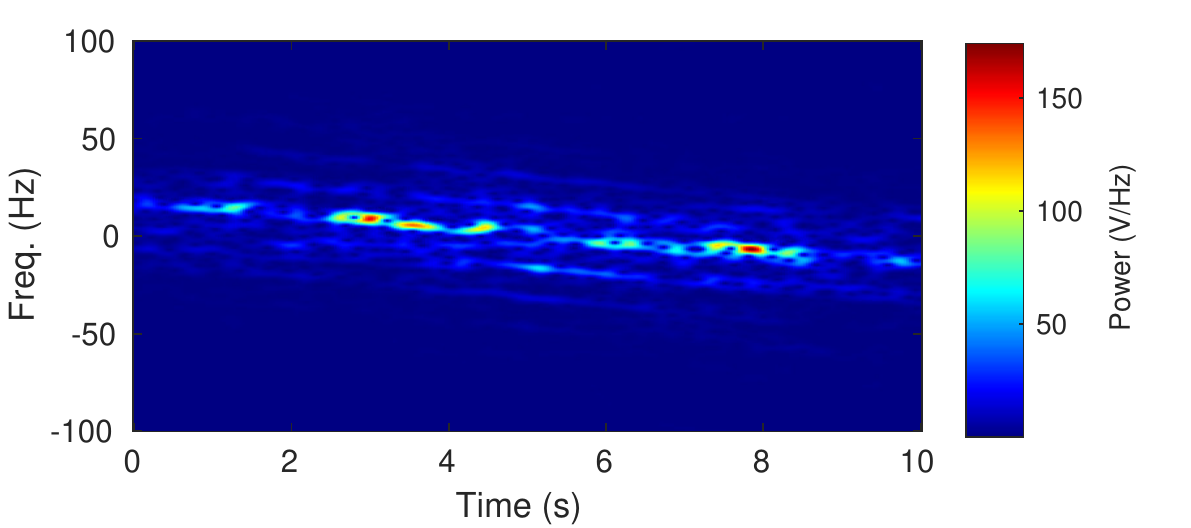}
\caption{}
\end{subfigure}
\begin{subfigure}{0.5\textwidth}
\includegraphics[width=\linewidth]{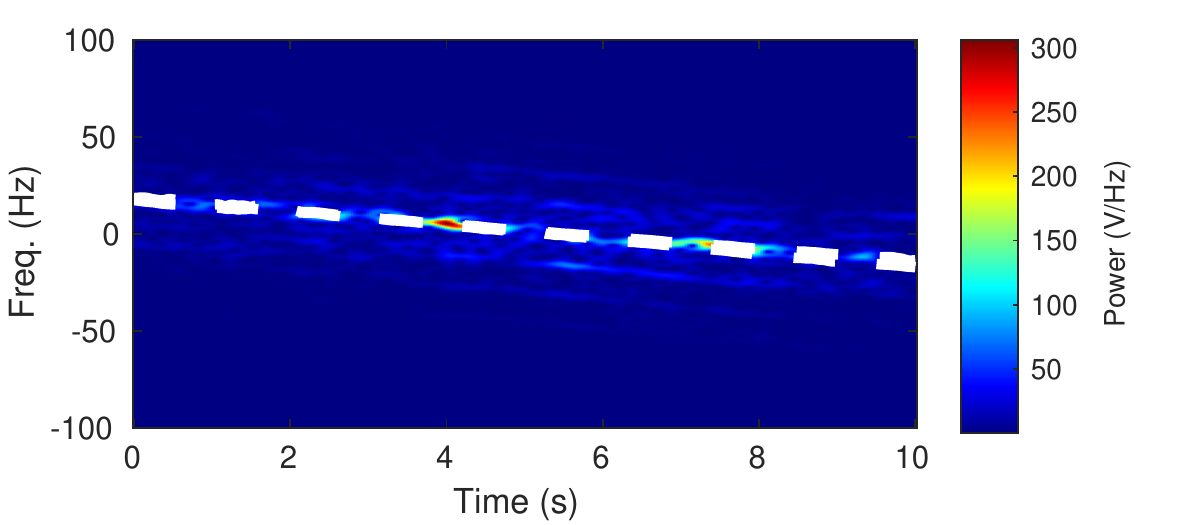}
\caption{}
\end{subfigure}
\begin{subfigure}{0.5\textwidth}
\includegraphics[width=\linewidth]{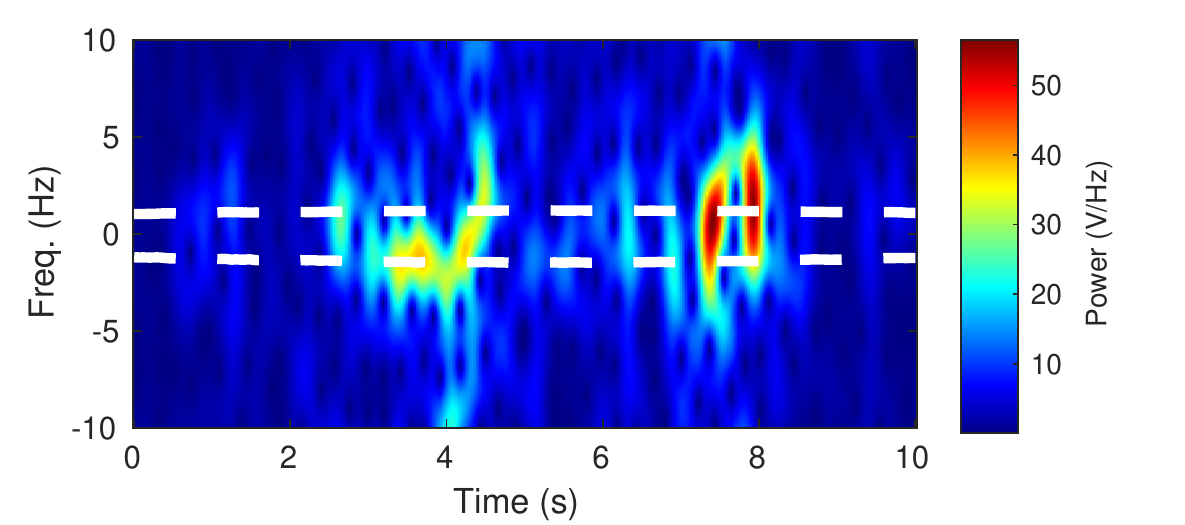}
\caption{}
\end{subfigure}
\caption{The measured time-frequency responses for the difficult case include the antenna one Doppler response (a), antenna two Doppler response with ground-truth expected Doppler shift (b), and interferometric response with the expected interferometric shift (c). 
}
\label{r42}
\end{figure}

The time-frequency for the first (easy) case is shown in \mbox{Fig. \ref{r37}}. The Doppler response for antenna one is shown (a), where we see large positive and negative Doppler shifts due to one target moving directly towards the radar and one moving away. The Doppler response from antenna two is shown (b) with the expected Doppler frequency shift overlaid in the white dashed line. The interferometric response is shown (c) also with the expected interferometric shift. Interestingly, in this case the distortion terms are at much higher frequencies, thus it may be inferred from the complete interferometric response that each target has no angular motion from the, even without the distortion mitigation technique. The time-frequency responses for the second (medium) case are shown in Fig. \ref{r40}. We see similar positive and negative frequency shifts due to the approaching and retreating targets which vary in time due to the angular motion. Here the interferometric response is corrupted by intermodulation distortion, and so estimating the individual target angular velocities from the complete interferometric response in (c) would be inaccurate. The time-frequency responses for the third (difficult) case are shown in Fig. \ref{r42}. Here we see the Doppler response of each target is identical and overlaid on top of each other. There is some interferometric response in the region of the expected interferometric shift, which is due to the fact that each target has the same Doppler shift and so the intermodulation distortion terms are driven to zero.

\subsection{Response Decomposition Angular Velocity Estimation}

Next we perform the response decomposition distortion mitigation technique, described above, on the measured responses, with a priori knowledge that $N=2$. We first present results assuming perfect knowledge of target location (in frequency) and perfect knowledge of target associations, obtained from the ground-truth system. After demonstrating the ideal case, we show practical results results the when target frequency locations must be detected and associated between antenna responses.

\subsubsection{Known Target Frequencies and Data Associations}

As shown in Fig \ref{r37} (b) we have access to the expected Doppler shift for each target, which is computed from the time-rate of change of the ground-truth position measurements. We then decompose each antenna response into two individual responses, one for each target, by masking out the spectral response with a passband of 10 Hz around the expected Doppler shift. This results in a total of four responses, the antenna one, target one response, the antenna one, target two response, the antenna two, target one response, and the antenna two, target two response, which are then cross-correlated pairwise with only like terms. By performing the individual cross-correlations in this manner, generation of the cross terms is avoided. We use the same mask to filter both the antenna one and antenna two responses for an individual target, which is possible since the combination of the interferometric baseline ($20\lambda$), and the velocities of the targets results in small differences in the Doppler frequencies between the targets. Thus, as long as the mask width is wide enough, the individual target response in both antenna signals is captures. As the baseline grows and the target velocities increase, there may be an appreciable difference in the Doppler frequencies between antennas for an individual target. Thus, in general one should perform separate detection and masking for each antenna response and for each target. Furthermore, since the difference in Doppler frequencies is small, data association is accomplished directly since only like targets are present in each masked signal. 

\begin{figure}[t]
\centering
\begin{subfigure}{0.5\textwidth}
\includegraphics[width=\linewidth]{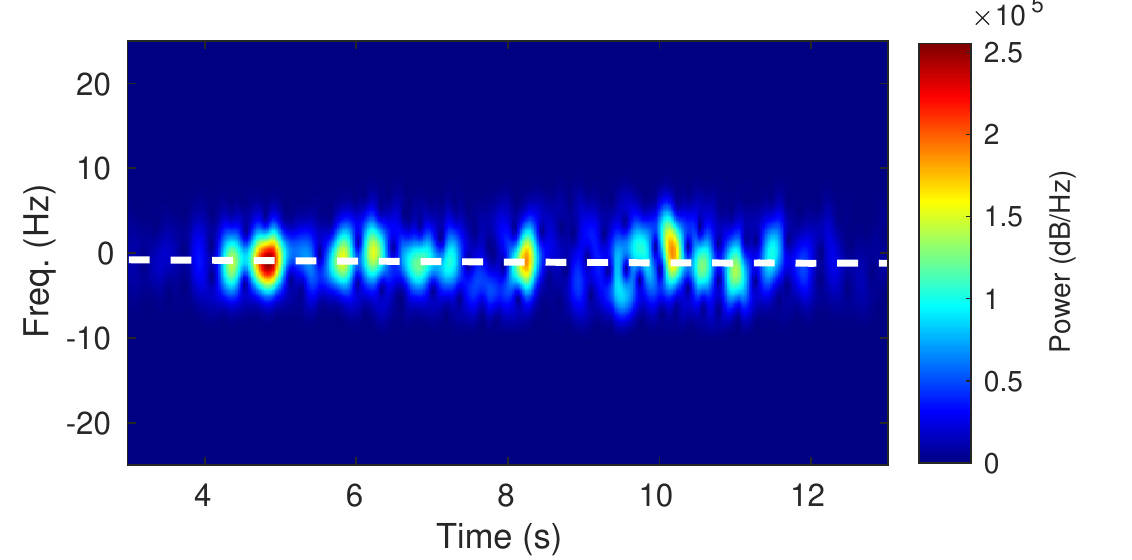}
\caption{}
\end{subfigure}
\begin{subfigure}{0.5\textwidth}
\includegraphics[width=\linewidth]{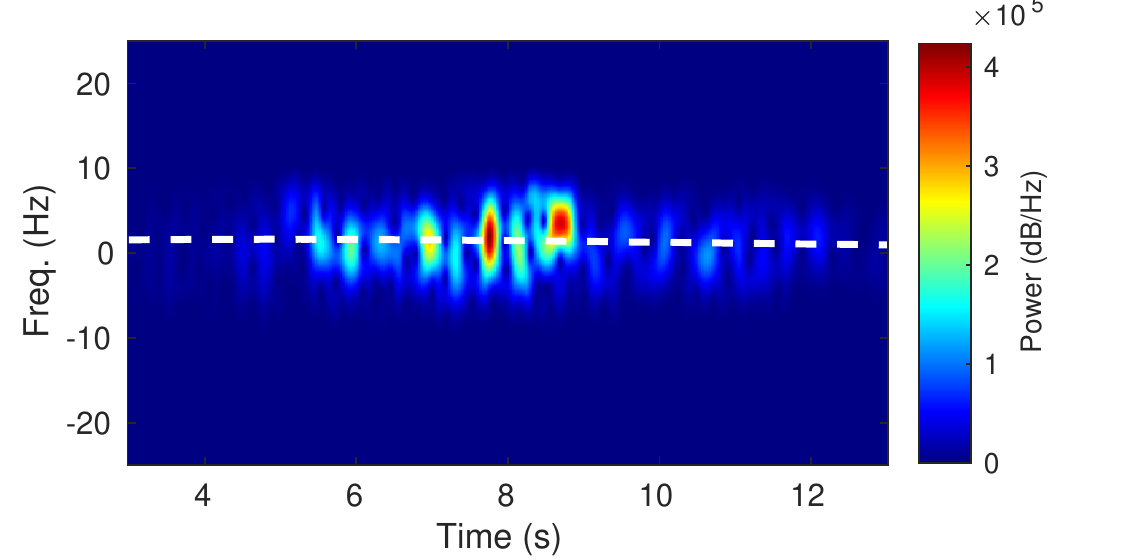}
\caption{}
\end{subfigure}
\caption{The reconstructed interferometric responses for the second case for target one (a), and target two (b). Here we have mitigated the intermodulation distortion which typically corrupts the complete interferometric response.
}
\label{interf_new}
\end{figure}

\begin{figure}[t]
\centering
\begin{subfigure}{0.5\textwidth}
\includegraphics[width=\linewidth]{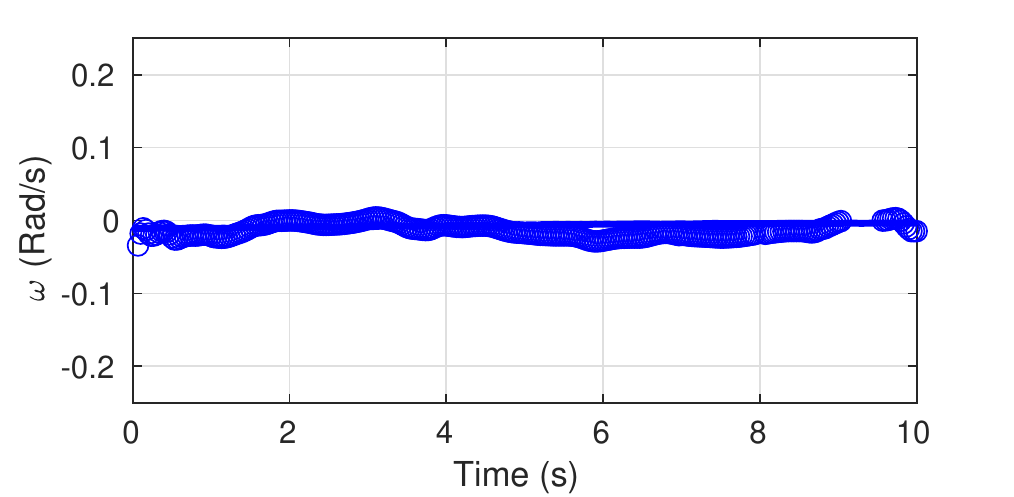}
\caption{}
\end{subfigure}
\begin{subfigure}{0.5\textwidth}
\includegraphics[width=\linewidth]{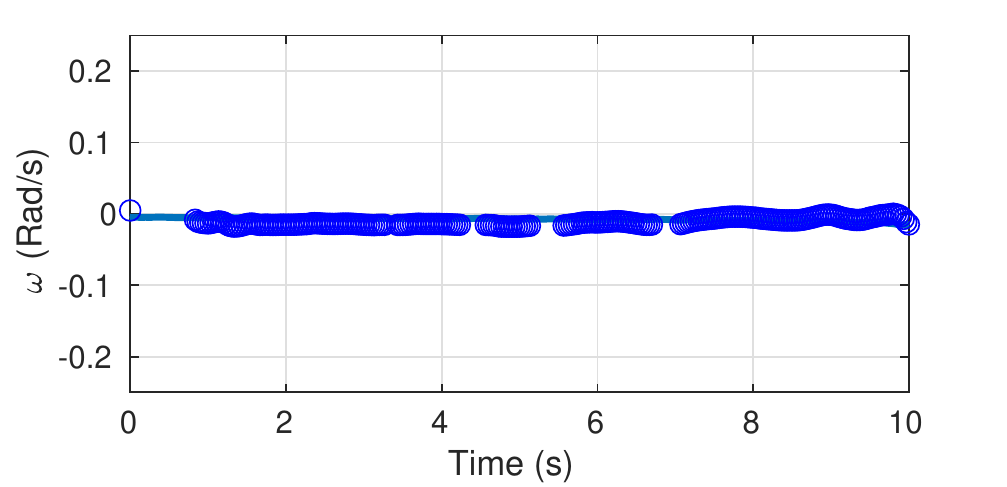}
\caption{}
\end{subfigure}
\caption{Angular velocity estimates for the first (easy) case with ground-truth angular velocity underlaid for target one (a) and target two (b). As expected we achieve accurate estimates with low variance in the simple case.
}
\label{r37_est}
\end{figure}

\begin{figure}[t]
\centering
\begin{subfigure}{0.5\textwidth}
\includegraphics[width=\linewidth]{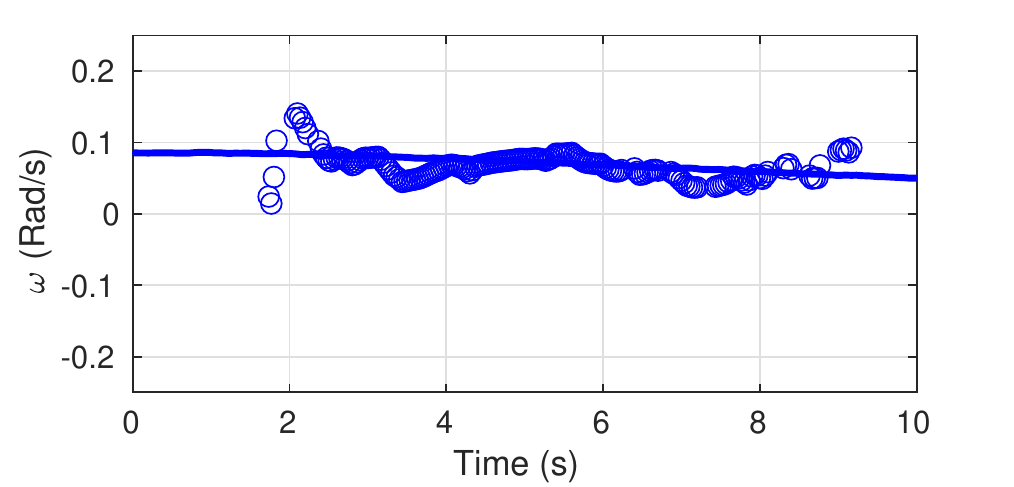}
\caption{}
\end{subfigure}
\begin{subfigure}{0.5\textwidth}
\includegraphics[width=\linewidth]{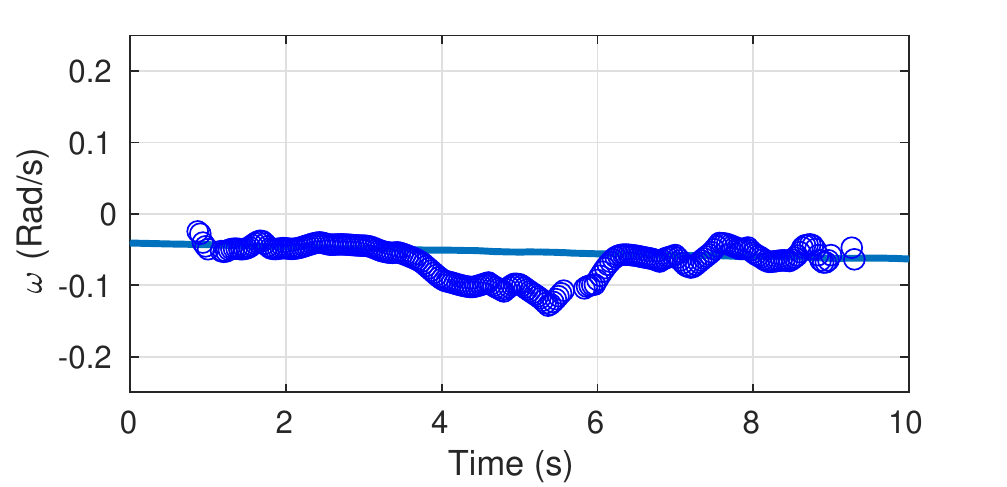}
\caption{}
\end{subfigure}
\caption{Angular velocity estimates for the second case with ground-truth angular velocity underlaid for target one (a) and target two (b). Here we see a higher variance than in the easy case, though still with relatively accurate estimates.}
\label{r40_est}
\end{figure}

\begin{figure}[t]
\centering
\begin{subfigure}{0.5\textwidth}
\includegraphics[width=\linewidth]{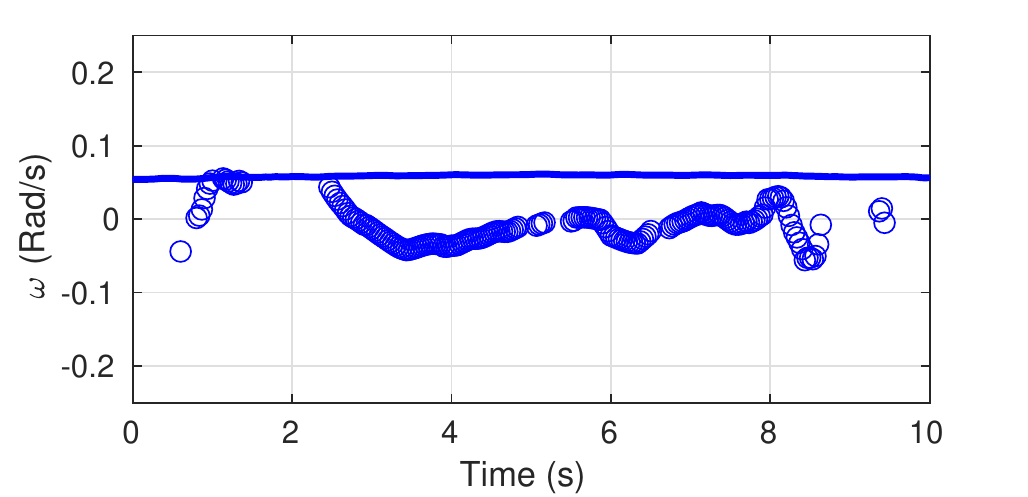}
\caption{}
\end{subfigure}
\begin{subfigure}{0.5\textwidth}
\includegraphics[width=\linewidth]{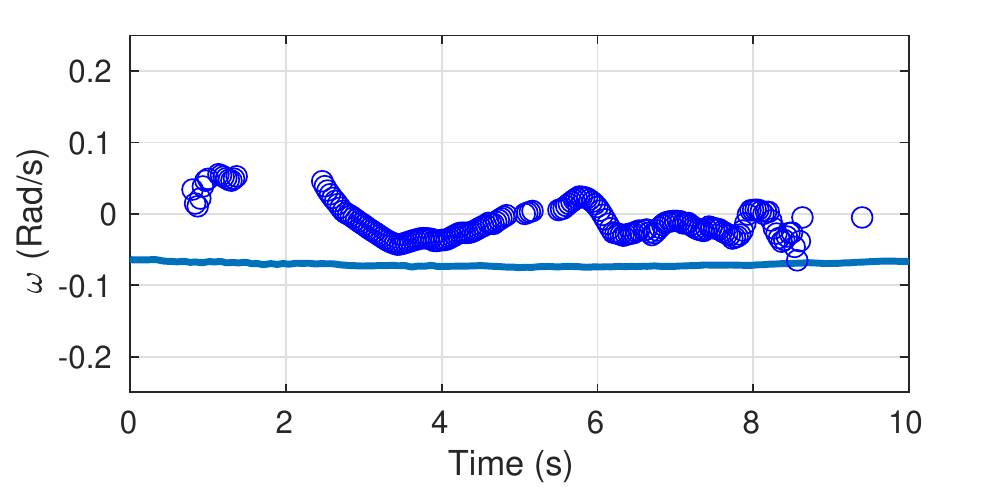}
\caption{}
\end{subfigure}
\caption{Angular velocity estimates for the third (difficult) case with ground-truth angular velocity underlaid for target one (a) and target two (b). As expected, due to the challenge of separating the Doppler responses, the estimates have a higher variance than the other cases.
}
\label{r42_est}
\end{figure}

After performing the cross-correlations in the frequency domain for each time instance we obtain two reconstructed interferometric responses, one for each target. These are shown for the second case in Fig. \ref{interf_new}, where we can identify the positive and negative interferometric frequency shift differences between each target, which clearly provides a more accurate estimate compared to the complete interferometric response with distortion shown in Fig. \ref{r40} (c). To estimate the interferometric frequency shift we take the maximum peak at each time instance of the reconstructed response with a cutoff threshold of -20 dB from the maximum peak voltage, which is used to eliminate areas with negligible signal response. We then perform smoothing with a window length of 60 samples or 0.5 s to reduce the variance of the estimates, reducing any noise due to platform vibrations and times where one target may be occluded. At each instant we use (\ref{shift}) with the small angle approximation to compute the angular velocities from the frequency estimates. The estimates for the first case are shown in Fig. \ref{r37_est} where, as expected, the estimated interferometric shift is close to zero since angular motion is negligible. The estimates for the second case are shown in Fig. \ref{r40_est}. Here we see an increase in both the estimation error and the variance, but the estimates are still within $30\%$ of the true values. Finally, the estimates for the third case are shown in Fig. \ref{r42_est}. As expected for the difficult case, the estimation error and variance increase appreciably. The estimation statistics for all cases are shown in Table \ref{est_stat}, where we compute the ground-truth mean angular velocity, $\mu$, the estimated mean from the response decomposition method, $\hat{\mu}$, and the standard deviation of the estimates.

\begin{table}[t!]
\caption{Angular Velocity Estimation Statistics With Known Target Frequencies}
\centering
\begin{tabular}{ccccc}
\hline\hline \\ [-2mm]
 \textbf{Target:}  & \textbf{$\mu$ (Rad/s):} & \textbf{$\hat{\mu}$ (Rad/s):} & \textbf{Std. (Rad/s)} \\ 
\hline \\ [-2mm]
Case 1, T1 & -0.007 & -0.014 & 0.009 \\
Case 1, T2 & -0.008 & -0.011 & 0.005 \\
Case 2, T1 & 0.072 & 0.067  & 0.019 \\
Case 2, T2 & -0.053 & -0.066  & 0.024 \\
Case 3, T1 & 0.059 & -0.008  & 0.025 \\
Case 3, T2 & -0.071 & -0.011  & 0.025 \\[0.5mm]
\hline         
\end{tabular}
\label{est_stat}
\end{table}

\subsubsection{Unknown Target Frequencies and Data Associations}

In the previous section we used the ground-truth system to locate and mask out the responses due to each target in the individual antenna responses, however this information may not always be available, thus the target responses must be detected and decomposed. In our measurements, we found that the response of an individual target did not resemble an ideal single frequency, but often had multiple peaks and harmonics which were associated with the vibration of the robotic base. Thus, instead of a simple peak detection, at each time instance we determine the two non-overlapping windows, each with a passband of 10 Hz, which had the highest total integrated power. Because the angular velocities are relatively low and the interferometric baseline is modest ($20\lambda$), the differences in target frequency between the two antenna responses is quite small. Because of this, we are able to use the same mask for each antenna response and so no data association is required. The estimation statistics for these cases are shown in Table \ref{est_stat2}. The results using this method were similar to the ideal scenario above, with only modestly higher estimation errors due to the uncertainty of the exact target locations. 


\begin{table}[t!]
\caption{Angular Velocity Estimation Statistics With Unknown Target Frequencies}
\centering
\begin{tabular}{ccccc}
\hline\hline \\ [-2mm]
 \textbf{Target:}  & \textbf{$\mu$ (Rad/s):} & \textbf{$\hat{\mu}$ (Rad/s):} & \textbf{Std. (Rad/s)} \\ 
\hline \\ [-2mm]
Case 1, T1 & -0.007 & -0.010 & 0.008 \\
Case 1, T2 & -0.008 & -0.008 & 0.004 \\
Case 2, T1 & 0.072 & 0.023  & 0.020 \\
Case 2, T2 & -0.053 & -0.036  & 0.015 \\
Case 3, T1 & 0.059 & -0.026 & 0.033 \\
Case 3, T2 & -0.071 & -0.001 & 0.019 \\[0.5mm]
\hline         
\end{tabular}
\label{est_stat2}
\end{table}





\section{Conclusion}

In this work we presented a new method of mitigating the impact of cross-term distortion in multitarget interferometric angular velocity estimation using a novel signal response decomposition method to avoid generating the distortion terms. By associating like target responses between each antenna signal we correlate each target only with itself, thus mitigating the cross-terms which are generated by performing the correlation in the time-domain. We presented angular velocity estimation results with a \mbox{40 GHz} interferometric radar for three target trajectory scenarios which were designed to highlight cases which were easy, medium, and difficult for the proposed distortion mitigation technique. Although we only present results for the simplest multitarget scenario of two targets, the presented response decomposition distortion mitigation method should scale as the number of targets increase.

\bibliographystyle{IEEEtran}
\bibliography{SP_distortion_mitigation_v5}

\vskip 0pt plus -1fil

\begin{IEEEbiography}[{\includegraphics[width=1in,height=1.25in,clip,keepaspectratio]{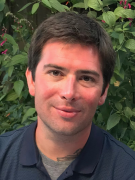}}]{Eric Klinefelter} received the B.S. degree in electrical engineering from Arizona State University with highest honors in 2018.  He is currently pursuing the Ph.D. degree in electrical engineering at Michigan State University, East Lansing, MI.  From 2018 to 2020 he was an active member of the MSU Autodrive team where he helped develop software used for low-level vehicle controls, waypoint following, HD mapping, and sensing.  In Spring 2019 he was a teaching assistant for the MSU course Creating Autonomous Vehicles, and in Summer 2019 was a lab instructor for an MIT Deeptech Bootcamp. His current research interests include millimeter-wave systems, machine learning, multiobject tracking, and autonomous vehicles. 
\end{IEEEbiography}
\vskip 0pt plus -1fil	

\begin{IEEEbiography}[{\includegraphics[width=1in,height=1.25in,clip,keepaspectratio]{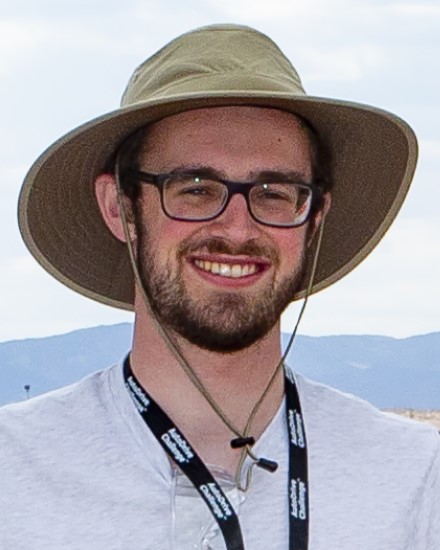}}]{Jason M. Merlo} received the B.S. degree in computer engineering from Michigan State University, East Lansing, MI, USA in 2018, where he is currently the Ph.D. degree in electrical engineering. His current research interests include distributed radar, wireless synchronization, interferometric arrays, and automotive radar applications.
\end{IEEEbiography}
\vskip 0pt plus -1fil

\begin{IEEEbiography}[{\includegraphics[width=1in,height=1.25in,clip,keepaspectratio]{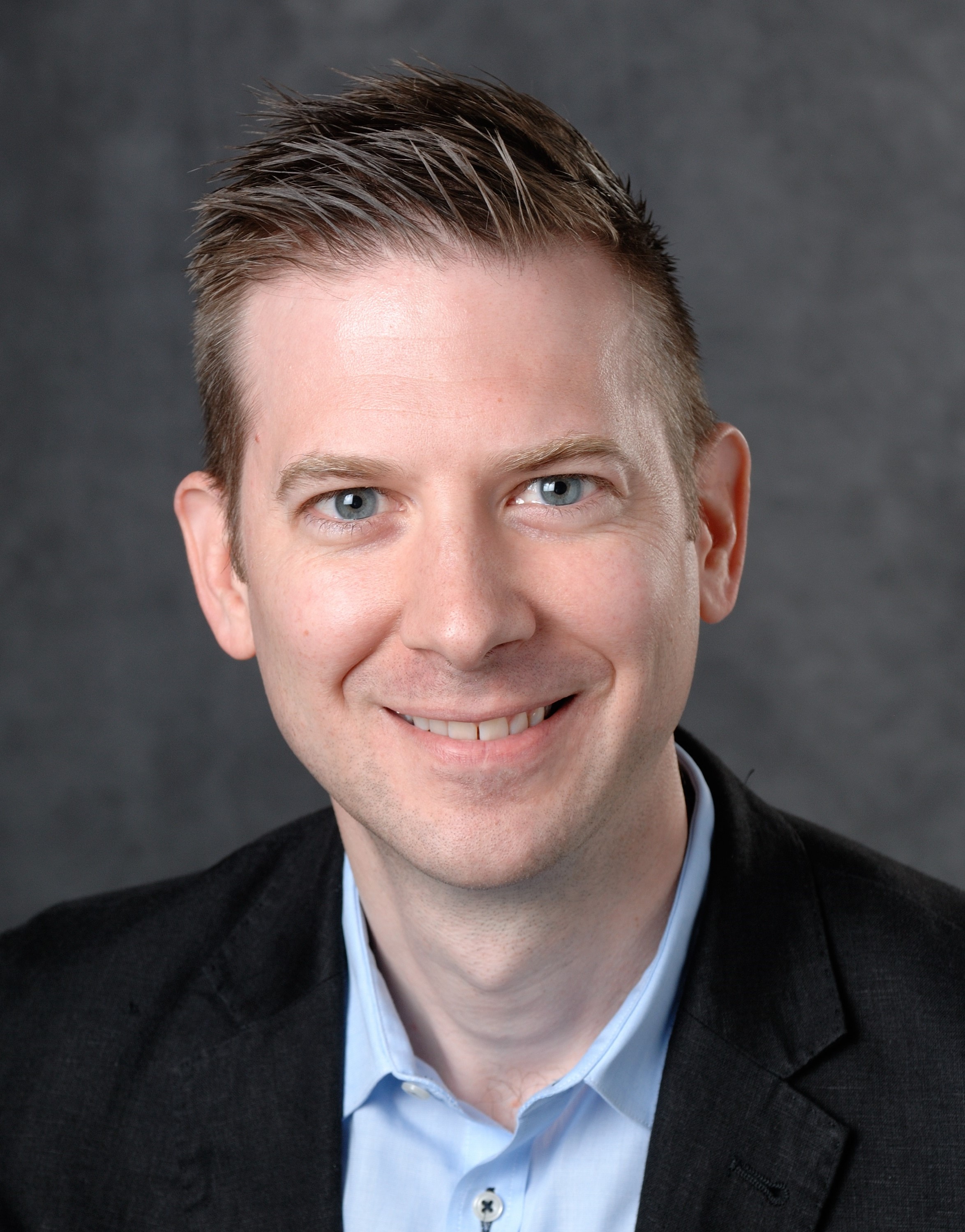}}]{Jeffrey A. Nanzer} (S'02--M'08--SM'14) received
	the B.S. degree in electrical engineering and the B.S. degree in
	computer engineering from Michigan State University,
	East Lansing, MI, USA, in 2003, and the M.S.
	and Ph.D. degrees in electrical engineering from The
	University of Texas at Austin, Austin, TX, USA,
	in 2005 and 2008, respectively.
	
	From 2008 to 2009, he was a Post-Doctoral Fellow
	with Applied Research Laboratories, University of
	Texas at Austin, where he was involved in designing
	electrically small HF antennas and communication
	systems. From 2009 to 2016, he was with The Johns Hopkins University
	Applied Physics Laboratory, Laurel, MD, USA, where he created and led
	the Advanced Microwave and Millimeter-Wave Technology Section. In 2016,
	he joined the Department of Electrical and Computer Engineering, Michigan
	State University, where he is currently the Dennis P. Nyquist Assistant
	Professor. He has authored or co-authored more than 175 refereed journal and
	conference papers, authored Microwave and Millimeter-Wave Remote Sensing
	for Security Applications (Artech House, 2012), and co-authored the chapter
	“Photonics-Enabled Millimeter-Wave Wireless Systems” in Wireless Transceiver
	Circuits (Taylor \& Francis, 2015). His current research interests include
	distributed arrays, radar and remote sensing, antennas, electromagnetics, and
	microwave photonics.
	
	Dr. Nanzer is a member of the IEEE Antennas and Propagation Society Education Committee and the USNC/URSI Commission B. He was a founding member and the First Treasurer of the IEEE APS/
	MTT-S Central Texas Chapter. He served as the Vice Chair for the IEEE Antenna Standards
	Committee from 2013 to 2015. He was the Chair of the Microwave Systems
	Technical Committee (MTT-16), IEEE Microwave Theory and Techniques
	Society from 2016 to 2018.
	He was a recipient of the Outstanding
	Young Engineer Award from the IEEE Microwave Theory and Techniques
	Society in 2019, the DARPA Director’s Fellowship in 2019, the National Science Foundation (NSF) CAREER Award in 2018, the DARPA Young
	Faculty Award in 2017, and the JHU/APL Outstanding Professional Book
	Award in 2012.  He is currently an Associate Editor of the IEEE
	TRANSACTIONS ON ANTENNAS AND PROPAGATION.
	
\end{IEEEbiography}

\end{document}